\documentclass[elsart,12pt]{article}
\usepackage{graphicx}
\usepackage{amssymb}
\usepackage{bm}
\newcommand{\hsp}{\hspace*{1pt}}
\newcommand{\bb}{\overline{B}}
\newcommand{\sbb}{\sum\limits_{j=B,\bb}}
\newcommand{\be}{\begin{equation}}
\newcommand{\ee}{\end{equation}}
\newcommand{\bel}[1]{\be\label{#1}}

\begin{document}
\vspace*{3cm}
\begin{center}
{\Large Clusters of matter and antimatter}\\
\vspace{0.5cm}
I.N. Mishustin\footnote{This work was supported in part by the Deutsche Forschung
Gemeinschaft (DFG) under grant 436 RUS 113/711/0-1, and the Russian Fond of 
Fundamental Research (RFFR) under grant 03-02-04007.} \\
\vspace{0.5cm}
\textit{Kurchatov Institute, Russian Research Center,
Moscow 123182, Russia}\\
\textit{Niels Bohr Institute, University of Copenhagen,
DK-2100 Copenhagen, Denmark}\\
\textit{Institut~f\"{u}r Theoretische Physik,
Goethe Universit\"{a}t, D--60054 Frankfurt am Main, Germany}
\end{center}


\begin{abstract}
In this talk I first give a short overview of antinuclei
produsction in recent experiments at RHIC.
Then I discuss the possibility of producing new types of nuclear
systems by implanting an antibaryon into ordinary nuclei. The structure
of nuclei containing one antiproton or antilambda is
investigated within the framework of a relativistic mean-field model.
Self-consistent calculations predict an enhanced binding and considerable
compression in such systems as compared with normal nuclei.
I present arguments that the life time of such nuclei with respect to
the antibaryon annihilation might be long enough for their observation.
Few experimental signatures are suggested. Possible formation of 
multi-quark-antiquark clusters is also discussed.

\end{abstract}

\section{Introduction}

Relativistic heavy-ion collisions open unique possibility of studying
antimatter production in the laboratory. Experiments with colliding Au
beams at highest energy available at present are conducted at the
Relativistic Heavy-Ion Collider (RHIC) in Brookhaven National
Laboratory. More than 1000 particles per unit rapidity are produced in
a central Au+Au collision at $\sqrt{s}=130$ AGeV. Most of them are
pions, but many baryons and antibaryons are produced too. Moreover, 
light nuclei and antinuclei have also been registered! As reported by the 
STAR experiment \cite{STAR}, the rapidity densities of different species 
at $y\approx 0$ are
\begin{equation}
\frac{dN}{dy}=32~({\rm
p}),~24~(\overline{\rm p}),~ 2.7\cdot 10^{-3}~(\overline{\rm d}),~
6.3\cdot 10^{-6}~(^3\overline{\rm He})~.
\end{equation}
No $^4\overline{\rm He}$ have been observed yet, although their
detection is feasible. The corresponding ratios are
\begin{equation}
\frac{\overline{\rm p}}{p}\simeq0.75,~~
\frac{\overline{\rm d}}{\overline{\rm p}}\simeq 10^{-4}, ~~
\frac{^3\overline{\rm He}}{\overline{\rm d}}\simeq 2\cdot 10^{-3}~.
\end{equation}
This is almost baryon-symmetric state of matter, similar to the one
existed in the early Universe around the deconfinment transition.

The BRAHMS experiment has measured rapidity densities of p and
$\overline{\rm p}$ in a wider rapidity interval, $0<y<3$ \cite{BRAHMS}.
Using these data one finds that the total number of antibaryons
produced in a central Au+Au collision is about 200. This is
sufficient to make an anti-Pb nucleus! But this is impossible
because these antibaryons are widely dispersed in the phase space.
Only lightest antinuclei ($\overline{\rm d}$, $\overline{\rm t}$,
$^{3,4}\overline{\rm He}$, ...) can be formed predominantly via the
coalescence mechanism.

One may speculate that more exotic clusters of baryons and
antibaryons, like $\overline{B}+B$, $\overline{B}+d$,
$\overline{B}+\alpha$ or even $\overline{d}+d$, can be formed in
ultrarelativistic heavy-ion collisions. If their life times are long
enough, they will show up themselves by the delayed annihilation.
More heavy clusters like $\overline{B}+A$ where $A$ stands for normal
nuclei ($^4$He, $^{16}$O, $^{40}$Ca,...) can be formed by using
high-energy antiproton beams. Below I discuss theoretical
predictions on properties of such clusters and their possible
signatures.  

\section{Relativistic description of nuclear matter}
It has been noticed already many years ago (see e.g. ref. \cite{Aue})
that nuclear physics may provide a unique laboratory for investigating
the Dirac picture of vacuum. The basis for this is given by
relativistic mean-field models which are widely used now for describing
nuclear matter and finite nuclei. Within this approach nucleons
are described by the Dirac equation coupled to scalar and vector
meson fields. Scalar $S$ and
vector $V$ potentials generated by these fields modify plane-wave
solutions of the Dirac equation as follows
\bel{pmnuc}
E^{\pm}({\bf p})=V\pm\sqrt{{\bf p}^2+(m-S)^2}~.
\ee
The $+$ sign corresponds to nucleons with positive energy
$E_N({\bf p})=E^{+}({\bf p})=V+\sqrt{{\bf p}^2+(m-S)^2}$~,
and the $-$ sign corresponds to antinucleons with energy
$E_{\bar{N}}({\bf p})=-E^{-}(-{\bf p})=-V+\sqrt{{\bf p}^2+(m-S)^2}$.
It is remarkable that changing sign of the vector potential for
antinucleons is exactly what is expected from the G-parity transformation
of the nucleon potential. As follows from eq. (\ref{pmnuc}),
the spectrum of single-particle states of the Dirac equation in nuclear
environment is modified in two ways.  First, the mass gap between
positive- and negative-energy states, $2(m-S)$, is reduced due to the
scalar potential and second, all states are shifted upwards due to the
vector potential.

It is well known from nuclear phenomenology
that good description of nuclear ground state is achieved with
$S\simeq 350$ MeV and $V\simeq 300$ MeV so that the net potential
for nucleons with $p\approx 0$ is $V-S\simeq -50$ MeV. Using the same values one
obtains for antinucleons a very deep potential, $-V-S\simeq -650$ MeV.
Such a potential would produce many strongly bound states in the Dirac
sea.  However, if these states are occupied they are hidden from
the direct observation.  Only creating a hole in this sea, i.e.
inserting a real antibaryon into the nucleus, would produce an
observable effect.  If this picture is correct one should expect the
existence of strongly bound states of antinucleons with nuclei \cite{Bue}.

The structure of such systems is calculated using several
versions of the relativistic mean--field model (RMF): TM1~\cite{sug94},
NL3 and NL-Z2~\cite{ben99}. Their parameters were found by fitting
binding energies and charge form-factors of spherical
nuclei from $^{16}$O to $^{208}$Pb. The general Lagrangian of the RMF
model is written as
\begin{eqnarray}
&&\hspace*{5mm}\mathcal{L}=\hspace*{-1mm}\sbb\overline{\psi}_j
\left(i\gamma^\mu\partial_\mu-m_j\right)\psi_j\nonumber\\
&&\hspace*{8mm}+\,\frac{1}{2}\hsp\partial^\mu\sigma\partial_\mu\sigma-
\frac{1}{2}\hsp m_\sigma^2\sigma^2-\frac{b}{3}\hsp\sigma^3-
\frac{c}{4}\hsp\sigma^4\nonumber\\
&&\hspace*{8mm}-\,\frac{1}{4}\hsp\omega^{\mu\nu}\omega_{\mu\nu}+
\frac{1}{2}\hsp m_\omega^2\omega^\mu\omega_\mu
+\frac{d}{4}\hsp (\omega^\mu\omega_\mu)^2\nonumber\\
&&\hspace*{8mm}-\,\frac{1}{4}\hsp\vec{\rho}^{\hsp\mu\nu}\vec{\rho}_{\mu\nu}+
\frac{1}{2}\hsp m_\rho^2\vec{\rho}^{\hsp\mu}\vec{\rho}_\mu \nonumber \\
&&\hspace*{8mm}+\hspace*{-1mm}\sbb\overline{\psi}_j\left(g_{\sigma j}\sigma+
g_{\omega j}\omega^\mu\gamma_\mu+
g_{\rho j}\vec{\rho}^{\hsp\mu}\gamma_\mu\vec{\tau}_j\right)\psi_j
\end{eqnarray}
plus Coulomb part. Here summation includes valence baryons $B$, i. e.
the nucleons forming a nucleus, and valence antibaryons $\overline{B}$
inserted in the nucleus. They are treated as Dirac particles coupled to
the scalar-isoscalar ($\sigma$), vector-isoscalar ($\omega$) and
vector-isovector ($\vec{\rho}$) meson fields. The calculations are
carried out within the mean-field approximation where the meson
fields are replaced by their expectation values. Also a "no-sea"
approximation is used. This implies that all occupied states of
the Dirac sea are "integrated out" so that they do not appear
explicitly.  It is assumed that their effect is taken into account by
nonlinear terms in the meson Lagrangian. Most calculations are done
with antibaryon coupling constants which are given by the G-parity
transformation ($g_{\sigma\overline{N}}=g_{\sigma N}$,
$g_{\omega\overline{N}}=-g_{\omega N}$) and $SU(3)$ flavor symmetry
($g_{\sigma\overline{\Lambda}}=\frac{2}{3}g_{\sigma\overline{N}}$,
$g_{\omega\overline{\Lambda}}=\frac{2}{3}g_{\omega\overline{N}}$).
In isosymmetric static systems the scalar and vector
potentials for nucleons are expressed as $S=g_{\sigma N}\sigma$
and $V=g_{\omega N}\omega^0$.
\begin{figure}
\vspace{-1cm}
\hspace*{-1cm}
\includegraphics[height=9cm]{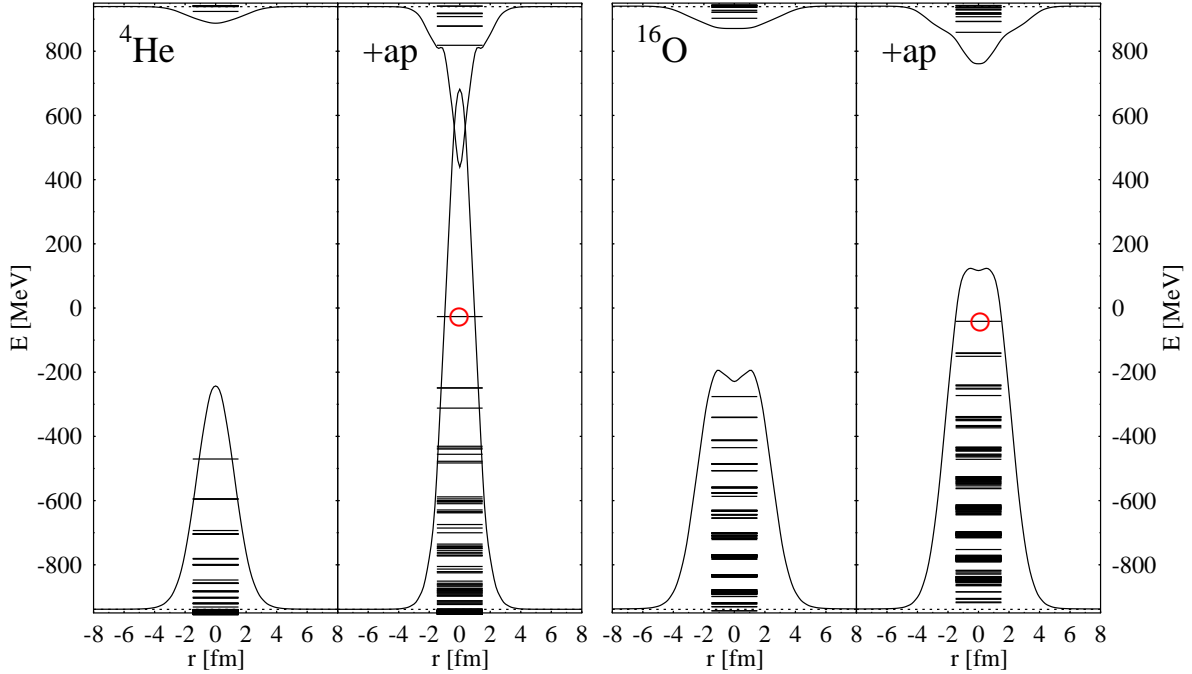}
\caption{
Single-particle levels in the upper and lower wells for $^4$He and
$^{16}$O nuclei calculated with the parametrization NL-Z2. Left
figures show ordinary nuclei with all levels in the lower well
occupied. Right figures show the same nuclei but containing one
antiproton (labelled as "+ap") i. e. one hole in the lower well.}
\label{fig1}
\end{figure}

\begin{figure}
\vspace*{-1cm}
\begin{center}
\hspace*{-0.5cm}
\includegraphics[height=16cm]{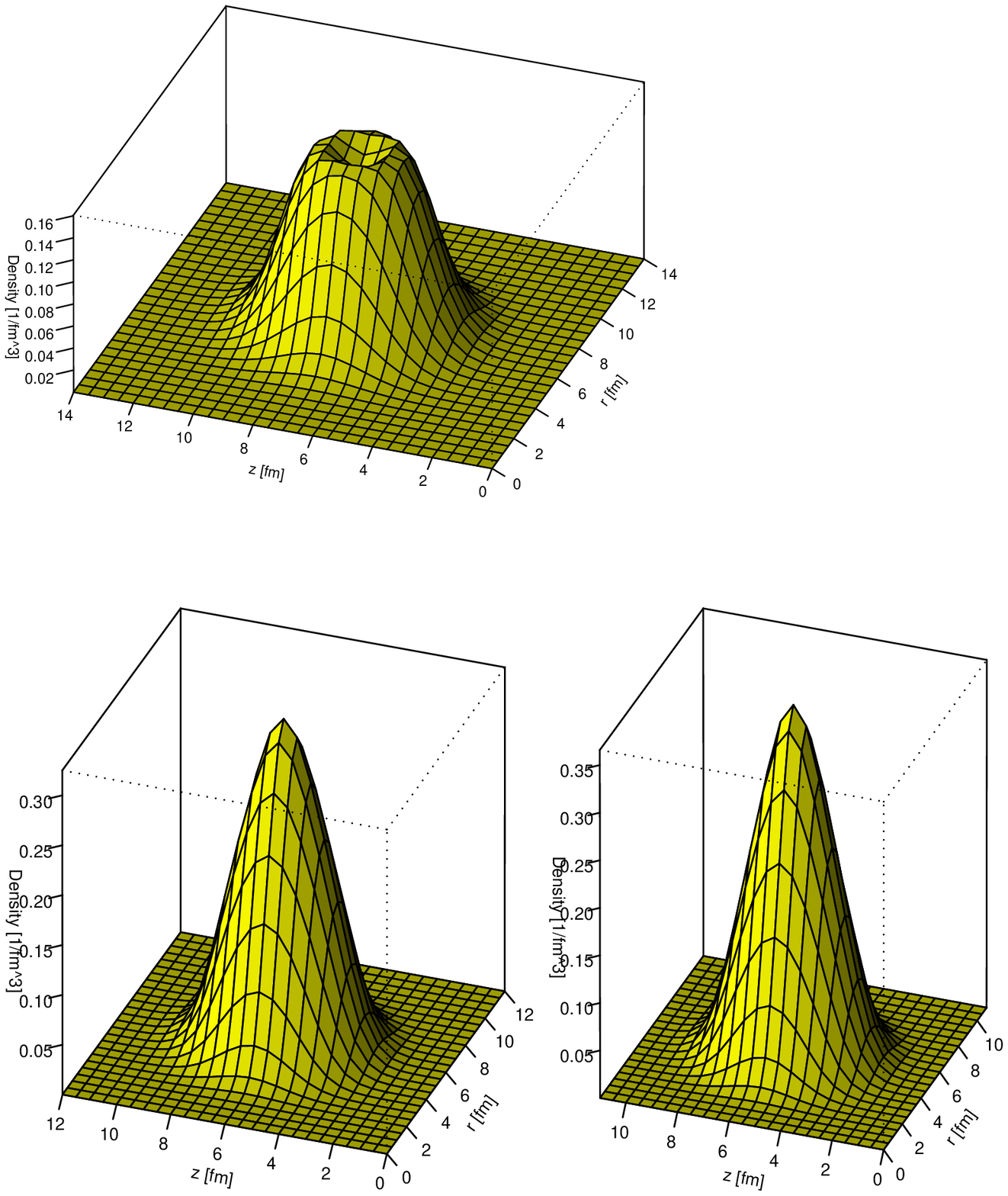}
\end{center}
\caption{
Sum of neutron and proton densities for $^{16}$O (top), $^{16}$O with
$\overline{p}$ (bottom right) and $^{16}$O with $\overline{\Lambda}$
(bottom left) calculated with the parametrization NL-Z2.}
\label{fig2}
\end{figure}

\section{Nuclear systems containing antibaryons}
Now I report on our recent study of antibaryon-doped nuclear systems
\cite{Bue}. Unlike some previous works, we take into account the
rearrangement of nuclear structure due to the presence of a real
antibaryon. Following the procedure suggested in Ref.~\cite{mao99}
and assuming the axial symmetry of the nuclear system, we solve
effective Schr{\"o}dinger equations for nucleons and
an antibaryon together with differential equations for mean
meson and Coulomb fields. We explicitly take into account
the antibaryon contributions to the scalar and vector densities.
It is important that antibaryons give a negative contribution to the
vector density, while a positive contribution to the scalar density.
This leads to increased attraction and decreased repulsion for surrounding
nucleons. To maximize attraction,
nucleons move to the center of the nucleus, where the antibaryon
has its largest occupation probability. This gives rise to a
strong local compression of the nucleus and leads to a dramatic
rearrangement of its structure.

\begin{figure}
\includegraphics[width=16cm]{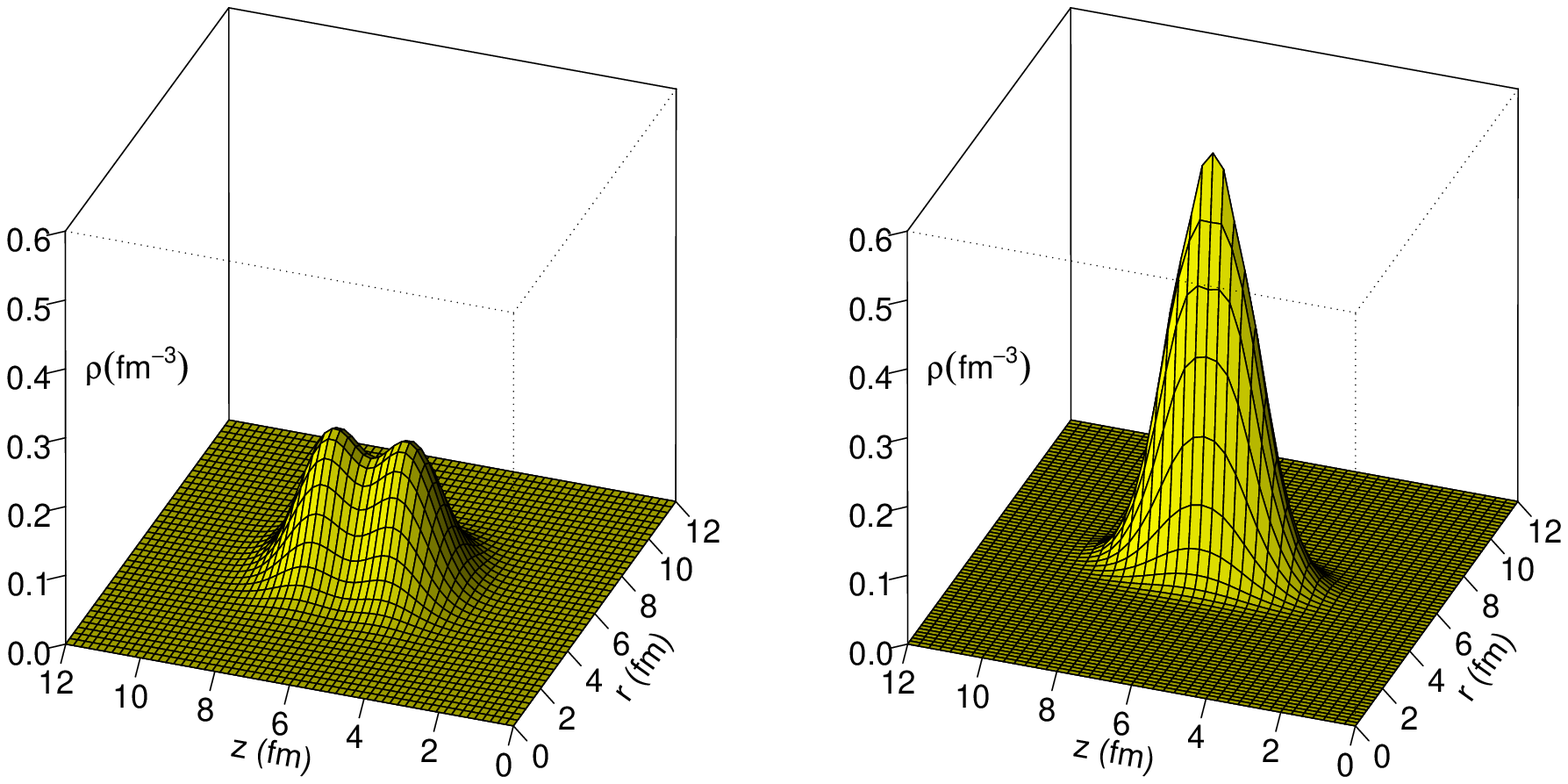}
\caption{
Nucleon densities for $^8$Be without (left) and
with (right) an antiproton calculated with
the parametrization NL3.}
\label{fig3}
\end{figure}

These features are illustrated by several examples
presented in Figs.~1-3. Figure 1 shows single-particle states in upper
and lower wells for $^4$He and $^{16}$O nuclei with and without an
antiproton. One can clearly see the rearrangement of nuclear structure
due to the presence of a hole in the Dirac sea. It is interesting to
note that the highest hole level (lowest antiproton level) moves to
about zero energy. This means that the whole antiproton rest mass is
eaten up by the attractive potential.

Figure 2 shows 3d plots of nucleon density distributions for $^{16}$O
nucleus with and without an antibaryon ($\overline{p}$,
$\overline{\Lambda}$) in a lowest energy state.  The calculations show
that inserting an antiproton into the $^{16}$O nucleus leads to the
increase of central nucleon density by a factor 2--4 depending on the
RMF parametrization. Due to a very deep antiproton potential the binding
energy of the whole system is increased significantly as compared with
130 MeV for normal $^{16}$O.  The calculated binding energies of the
$\overline{p}+^{16}$O system are 830, 1050 and 1160 MeV for the NL--Z2,
NL3 and TM1, respectively.  Due to this anomalous binding we call such
systems super bound nuclei (SBN).  In the case of antilambdas we
rescale the coupling constants with a factor 2/3 that leads to the
binding energy of 560$\div$700 MeV for the $\overline{\Lambda}+^{16}$O
system.

As second example, we investigate the effect of a single antiproton
inserted into the $^8$Be nucleus.
The normal $^8$Be  nucleus is not spherical,
exhibiting a clearly visible $2\alpha$~structure with the
ground state deformation $\beta_2\simeq 1.20$\,.
As seen in Fig.~3, inserting an
antiproton in $^8$Be results in a much less elongated shape
($\beta_2\simeq 0.23$) and disappearance of its cluster structure.
The binding energy increases from $53$~MeV to about $700$~MeV.
Similar, but weaker effects have been predicted~\cite{aka02}
for the $K^-$ bound state in the $^8$Be nucleus.

The calculations have been performed also with reduced antinucleon coupling
constants as compared to the G-parity prescription. We have found that
the main conclusions about enhanced binding and considerable compression
of $\overline{p}$-doped nuclei remain valid even when
these coupling constants are reduced by factor 3 or so.

\section{Antibaryon annihilation in nuclei}
The crucial question concerning possible observation of the SBNs
is their life time. The main decay
channel for such states is the annihilation of antibaryons on
surrounding nucleons.
The energy available for annihilation of a bound antinucleon
equals \mbox{$Q=2m_N-B_N-B_{\overline{N}}$},
where $B_N$ and $B_{\overline{N}}$ are the corresponding binding energies.
In our case this energy is at least by a factor 2 smaller
as compared with the vacuum value of $2m_N$. This should lead to
a significant suppression of the available phase space and thus to
a reduced annihilation rate in medium. We have performed detailed
calculations assuming that the annihilation rates into different
channels are proportional to the available phase space. All
intermediate states with heavy mesons like $\rho$, $\omega$, $\eta$
as well as multi-pion channels have been considered. Our conclusion is that
decreasing the Q value from 2 GeV to 1 GeV may lead to the reduction of total
annihilation rate by factor 20$\div$30.	Then we estimate the SBN life
times on the level of 5-25 fm/c which makes their
observation feasible. This large margin in the life times is mainly
caused by uncertainties in the overlap integral between antinucleon and
nucleon scalar densities. Longer life times may be expected for SBNs
containing antihyperons. The reason is that, instead of pions, more heavy 
particles, kaons, are requird for their annihilation.  We have
also analyzed multi-nucleon annihilation channels (Pontecorvo-like
reactions) and have found their contribution to be less than 40\% of
the single-nucleon annihilation.

\begin{figure}
\vspace{-1cm}
\hspace*{-1cm}
\includegraphics[height=8cm]{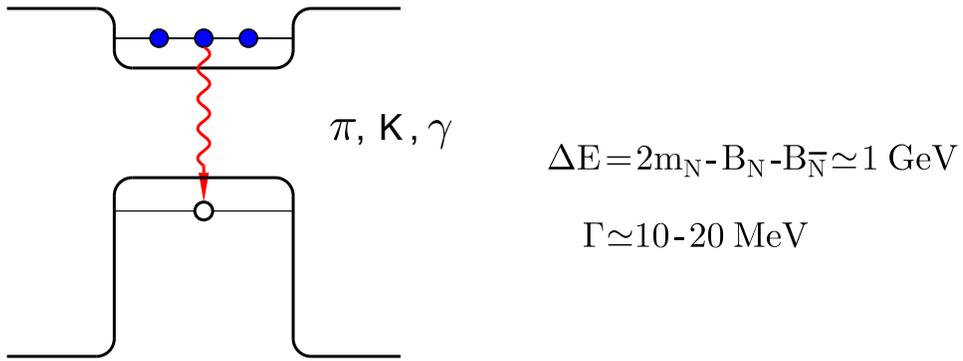}
\vspace{-1cm}
\caption{
One-body annihilation leading to sharp lines in the energy spectrum of
emitted particles.}
\label{fig4}
\end{figure}
\begin{figure}
\vspace*{-3cm}
\includegraphics[height=8cm]{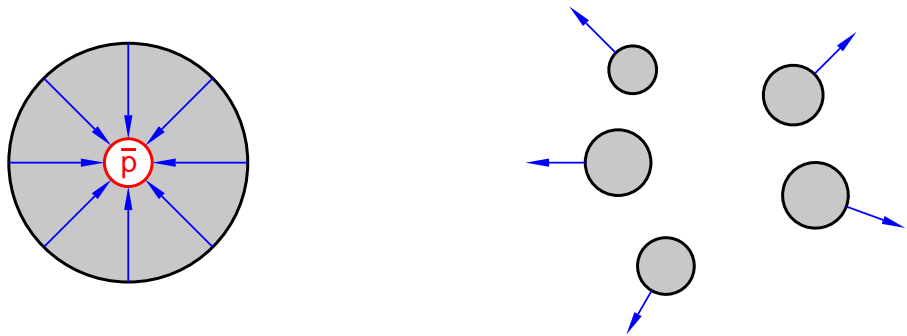}
\vspace*{1cm}
\caption{
Left: compression of a nucleus induced by an antibaryon. Right:
collective flow of nuclear fragments resulting from the expansion and
disintegration of the nucleus after antibaryon annihilation.}
\label{fig5}
\end{figure}

\section{Formation and signatures of SBNs}
We believe that such exotic antibaryon-nuclear systems can be produced
by using antiproton beams of multi-GeV energy, e.g. at the future GSI
facility. It is well known that low-energy antiprotons annihilate on
the nuclear periphery (at about 5$\%$ of the normal density).  Since
the annihilation cross section drops significantly with energy, a
high-energy antiproton can penetrate deeper into the nuclear interior.
Then it can be stopped there in an inelastic collision with a nucleon,
e. g. via the reaction
$A\hspace{1pt}(\bar{p},N\pi)\hspace{1pt}_{\bar{p}}A'$, leading to the
formation of a $\bar{p}$-doped nucleus. Reactions like
$A\hspace{1pt}(\bar{p},\Lambda)\hspace{1pt}_{\overline{\Lambda}}A'$
can be used to produce a $\overline{\Lambda}$-doped nuclei.
Fast nucleons or lambdas can be used for triggering such events.
In order to be captured by a target-nucleus final antibaryons
must be slow in the lab frame.
Rough estimates of the SBN formation probability in a central
$\bar{p}A$ collision give the values \mbox{$10^{-5}-10^{-6}$}. With the
$\bar{p}$ beam luminocity of 2$\cdot 10^{32}$ cm$^{-2}$s$^{-1}$
planned at GSI this will correspond to the reaction rate from tens to 
hundreds desired events per second.

Several signatures of SBNs can be
used for their experimental observation. First,
annihilation of a bound antibaryon can proceed via
emission of a single photon, pion or kaon with an energy of about
1~GeV (such annihilation channels are forbidden
in vacuum). So one may search for relatively sharp lines,
with width of 10$\div$40 MeV, around this
energy, emitted isotropically in the SBN rest frame (see Fig.~4).
Another signal may come from explosive disintegration of the
compressed nucleus after the antibaryon annihilation (see Fig.~5).
This process can be observed by measuring radial collective velocities
of nuclear fragments.

\section{Multi-quark-antiquark clusters}
It is interesting to look at the antibaryon-nuclear systems from
somewhat different point of view. An antibaryon
implanted into a nucleus acts as an attractor for surrounding nucleons.
Due to the uncompensated attractive force these nucleons acquire
acceleration towards the center. As the result of this inward collective
motion the nucleons pile up producing local compression.
If this process would be completely elastic it would generate monopole-like
oscillations around the compressed SBN state.
\begin{figure}
\vspace{-2cm}
\hspace*{-1cm}
\includegraphics[height=10cm]{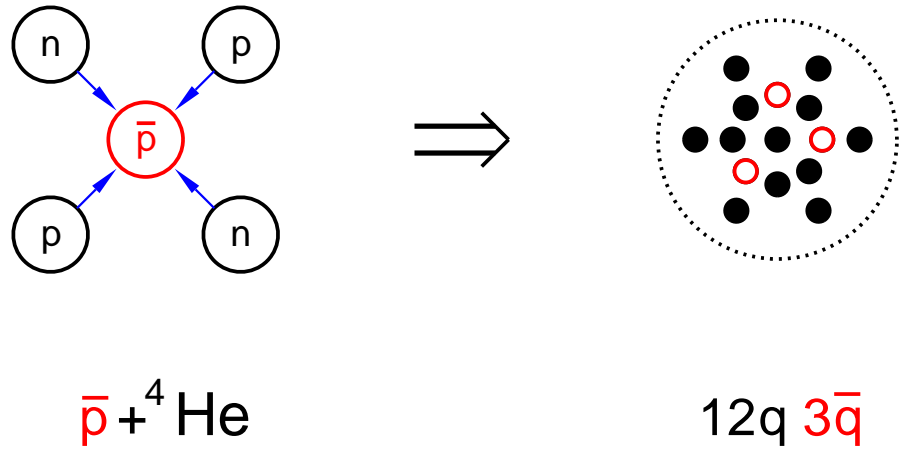}
\caption{
Formation of a multi-quark-antiquark cluster at an intermediate stage
of the antibaryon annihilation in a $^4$He nucleus.}
\label{fig6}
\end{figure}
The maximum compression is
reached when the attractive potential energy becomes equal to the
compression energy. Simple estimates show that local baryon densities
up to 5 times the normal nuclear density may be obtained in this way.
It is most likely that the deconfinment transition will occur at
this stage and a high-density cloud containing an antibaryon and a few
nucleons will appear in the form of a multi-quark-antiquark
cluster. One may speculate that the whole $^4$He or even $^{16}$O nucleus can
be transformed into the quark phase by this mechanism. This process is
illustrated in Fig.~6.
As shown in ref.  \cite{Mis1}, an admixture of
antiquarks to cold quark matter is energetically favorable. The problem
of annihilation is now transferred to the quark level.  But the
argument concerning the reduction of available phase space due to the
entrance-channel nuclear effects should work in this case too.  Thus
one may hope to produce relatively cold droplets of the quark phase by
the inertial compression of nuclear matter initiated by an antibaryon.

I am grateful to T. B\"urvenich, W. Greiner and L.M. Satarov
for fruitful discussions and help in the preparation of this talk.

\end{document}